\documentstyle[11pt,aaspp4]{article}
\slugcomment{\shortstack[r]{KUNS-1640}}
\def\lsim{\mathrel{\mathpalette\Oversim<}}
\def\gsim{\mathrel{\mathpalette\Oversim>}}
\def\Oversim#1#2{\lower0.5ex\vbox{\baselineskip0pt\lineskip0pt%
            \lineskiplimit0pt\ialign{%
          $\mathsurround0pt #1\hfil##\hfil$\crcr#2\crcr\sim\crcr}}}

\begin{document}
\def\be{\begin{equation}}
\def\ee{\end{equation}}
\def\ba{\begin{eqnarray}}
\def\ea{\end{eqnarray}}

\title{Self-Regulation of Star Formation in Low Metallicity Clouds}
\author{Ryoichi Nishi 
and 
Motomichi Tashiro
\affil{Department of Physics, Kyoto University, Kyoto 606-8502, Japan}}

\abstract{
We investigate the process of self-regulated  
star formation via photodissociation of 
hydrogen molecules in low metallicity clouds. 
We evaluate the influence region's scale of a massive  
star in low metallicity gas clouds whose temperatures are between 
$10^2$ and $10^4$ Kelvin. 
A single O star can photodissociate H$_2$ in 
the whole of the host cloud. 
If metallicity is smaller than about 
$10^{-2.5}$ of the solar metallicity, 
the depletion of coolant of the the host cloud is very 
serious so that the cloud cannot cool in a
free-fall time, and subsequent star formation is almost quenched.
On the contrary, if metallicity is larger than about 
$10^{-1.5}$ of the solar metallicity, 
star formation regulation via photodissociation is not efficient. 
The typical metallicity when this transition occurs is  
$\sim 10^{-2}$ of the solar metallicity.   
This indicates that stars do not form efficiently 
before the metallicity becomes larger than 
about $10^{-2}$ of the solar metallicity and we considered 
that this value becomes the lower limit of the metallicity 
of luminous objects such as galaxies. 
}

\keywords{cosmology: theory --- early universe --- galaxies: formation 
--- H II region --- ISM: clouds --- stars: formation}

\newpage
\section{Introduction} 

After the recombination era, little information is accessible 
until $z \sim 5$, 
after that we can observe objects such as galaxies and QSOs. 
On the other hand, the reionization of the intergalactic medium and 
the presence of heavy elements at high-$z$  
suggest that there are other populations of luminous objects, 
which precedes normal galaxies.
Thus, a theoretical approach to reveal the formation mechanism of 
such unseen luminous objects is very important.

The formation process of a luminous object is roughly divided into 
three steps, formation of cold clouds by H and/or ${\rm H_2}$ line 
cooling, formation of the first generation stars in the cold clouds, 
and the star formation throughout the clouds. However, the mass 
of the first generation stars are estimated through detailed 
investigation to be fairly large (\cite{NU99,ON98}). 
Thus, the third step is  
disturbed by the feedback from the massive stars formed in the clouds. 
The main feedback consists of two 
different processes, UV radiation from the stars and energy input by 
supernovae (SNe). 
Through ionization of H (\cite{LM92}) and dissociation of H$_2$ 
(\cite{Silk77,ON99}), ultra violet (UV) radiation has negative feedback 
on further star formation in the host clouds. 
Especially, H$_2$ is dissociated in such 
a large region that the whole of ordinary low mass cosmological object 
is influenced by one O5 type star (\cite{ON99}). 
For the case of a metal-free gas cloud, the influence region is much 
wider than the HII region and the metal-free host cloud lacks coolant 
and cannot cool. 
Thus, next generation stars are hardly formed before the first generation 
stars die. 
However, the life time of massive stars is much shorter than 
the cosmological time scale and they die as SNe. 
By these SN explosions, the cloud's gas is often dispersed  
before significant amount 
of total gas is transformed in stars (e.g., \cite{MF99,Ciardi99,NS99}). 
On the other hand, if the gas binding is not disrupted, 
next generation stars are formed in a cloud which is slightly polluted 
by heavy elements. 
Even in the case that the host cloud is disrupted by SN explosions, 
if the remnant gas does not escape from the host pregalactic object, 
next generation clouds, which is slightly polluted, will be formed 
and subsequent star formation will follow. 
In these polluted clouds, heavy elements will become important coolants, 
if their abundances increase to some degree. 
After the host cloud is enough polluted that star formation regulation 
by UV radiation is not efficient, the effective star formation 
can start. Thus, the pregalactic object, which is a cloud complex 
will evolve into a luminous object such as galaxy. 

In this paper, 
we investigate the self regulation of star formation via UV radiation, 
and assess the critical metallicity which enables the formation 
of luminous objects. 

\section{Influence Region of a Massive Star in a Low Metallicity Cloud}

Around an OB star, hydrogen is photoionized, and an HII region is formed. 
Lin \& Murray (1992) considered the star formation regulation via 
photoionization. However, the regulation 
can be efficient outside the HII region via photodissociation of H$_2$ 
in a low metallicity cloud, where H$_2$ line emissions are the most 
important coolant. 

Although ionizing photons hardly escape from the HII region, photons whose
radiation energy are below the Lyman limit can get away.
Such UV photons photodissociate H$_2$, and a 
photodissociation region (PDR) is formed around the HII region. 
In a PDR in a metal-free cloud, H$_2$ dissociation effect is very 
efficient so that the region which is larger than 
the whole of cosmological low mass cloud  
is influenced by only one O5 type star (\cite{ON99}). 
However, after a cloud is polluted by heavy elements, 
the situation becomes complicated, since other thermal processes may 
be important in a PDR in a cloud with heavy elements. 
In the region,  
CO molecules are dissociated also, since the threshold UV energy of 
H$_2$ and CO dissociation are close, and  
C, Si and Fe in the gas phase are ionized, 
since ionizing energies of C, Si and Fe are lower than H. 
Thus, C$^+$, Si$^+$ and Fe$^+$ cooperate with H$_2$ as main coolants in a 
low metallicity cloud. 
On the other hand, dust photoelectric heating becomes important heating 
source in a polluted cloud. 
In this section, we study how much mass in a low metallicity cloud is
affected by UV photons from an OB star and, as a result, becomes
unable to cool in a free-fall time. 

To calculate the heating rate ($\Gamma$) and the cooling rate ($\Lambda$) 
per unit volume, we use the rates of Wolfire et al. (1995) 
and the references there in, 
and the rates of Galli \& Palla (1999) (processes related to H$_2$). 
But we do not include the effects of X-ray and cosmic ray. 
We assume the ionization degree $x_e$ as 
\be 
x_e = x_{\rm H^+} + x_{\rm C^+} + x_{\rm Si^+} + x_{\rm Fe^+},
\ee
where $x_i$ is the abundance of the $i$ element. 
We assume that abundances of heavy elements are determined from 
the cosmic abundances by scaling proportional to 
$z/z_\odot$\footnote{The
depletion of the gas phase abundance of heavy elements is serious 
in the interstellar clouds. 
However, the main coolant except H$_2$ is C$^+$ and O, and the depletion of 
C and O is not so large. Moreover, the depletion of C may not be serious 
considering dust formation in SN ejecta (e.g., \cite{Kozasa89}).
But it may be possible that the depletion of C and O is more significant 
than the above estimate. 
In this case,  
we should consider that the gas phase abundances of C and O 
represent the heavy element 
abundance, and hence main results in this 
paper are almost the same.}. 
The adoptive values are $x_{\rm C} = 10^{-3.52} z/z_\odot$, 
$x_{\rm O} = 10^{-3.34}z/z_\odot$, $x_{\rm Si} = 10^{-5.45}z/z_\odot$ 
and $x_{\rm Fe} = 10^{-6.15}z/z_\odot$. We assume that all of these 
elements are ionized.  
Here we add the extra term $x_{\rm H^+}$ to evaluate  
the effect of relic ionization 
of cosmological recombination and/or previous SN, etc.. 
We investigate for the cases of $x_{\rm H^+} = 10^{-4}$ and 
$x_{\rm H^+} = 0$. However, the overall tendency is not affected 
by the value of $x_{\rm H^+}$. Thus, hereafter, we show the results 
for the case of $x_{\rm H^+} = 10^{-4}$ mainly. 
In a PDR, H$_{2}$ molecules are dissociated mainly via 
the two-step photodissociation
process by the Lyman and Werner (LW) bands photons. 
For H$_2$ number density, we use the equilibrium
value with the initial ionization degree (\cite{ON99}).  
This treatment may result in overestimation of $x_{\rm H_2}$ hence in 
overestimation of cooling rate (see e.g., \cite{NS99}). 
However, to seek the lower bound of the region affected by the 
photodissociating UV radiation from a massive star, we use the 
equilibrium value. 

H$_{2}$ is formed mainly via the H$^{-}$ process
\begin{eqnarray}
\label{eq:Hm1}
{\rm H}+e^{-} &\rightarrow & {\rm H^{-}}+\gamma  \\
\label{eq:Hm2}
{\rm H}+{\rm H^{-}} &\rightarrow & {\rm H_{2}}+e^{-}.
\end{eqnarray}
The rate-determining stage of the H$^{-}$ process is the reaction
 (\ref{eq:Hm1}), whose rate coefficient $k_{\rm H^{-}}$  is (\cite{dJ72})
\begin{equation}
  k_{\rm H^{-}}=1.0 \times 10^{-18} T~{\rm s^{-1} cm^{3}}. 
\end{equation}
In a PDR, H$_{2}$ is dissociated mainly via the two-step photodissociation
process 
\begin{equation}
  {\rm H_2}+\gamma  \rightarrow {\rm H_{2}^*} \rightarrow 2{\rm H},
\end{equation}
where rate coefficient $k_{\rm 2step}$ is given by 
(\cite{KBS97,DB96})
\begin{equation}
  k_{\rm 2step}=1.13 \times 10^{8} F_{\rm LW}~{\rm s^{-1}}. 
\end{equation}
Here $F_{\rm LW}~({\rm ergs~s^{-1}cm^{-2}Hz^{-1}})$ is the averaged
radiation flux in the Lyman and Werner (LW) bands. 
Thus, the equilibrium number density of H$_{2}$ under ionization degree 
$x_e$ is
\begin{eqnarray}
\label{eq:neq}
n_{\rm H_2} &=& \frac{k_{\rm H^{-}}}{k_{\rm 2step}}x_e n_{\rm H}^{2} \cr
&=& 0.88 \times 10^{-26} x_e F_{\rm LW}^{-1} T  n_{\rm H}^{2},  
\end{eqnarray}
where $n_{\rm H}$ is the number density of the Hydrogen nuclei. 
On the other hand, the averaged flux in 
the Lyman and Werner bands is approximately given by 
\begin{equation}
F_{\rm LW}=\frac{L_{\rm LW}}{4 \pi r^2}. 
\end{equation}
Then, if self shielding effect can be neglected, 
$n_{\rm H_2}$ is proportional to $r^2$ and only at the very outer
region, $x_{\rm H_2}$ becomes abundant enough for efficient cooling. 

However, self shielding effect is important for ordinary clouds 
considered in this paper. 
If column density of H$_2$ becomes larger than $10^{14}~{\rm cm}^{-2}$, 
$F_{\rm LW}$ decreases because of self shielding (\cite{DB96}). 
At the outer region where LW band radiation is shielded, 
H$_{2}$ is dissociated via thermal collision and $n_{\rm H_2}$ becomes 
thermal equilibrium value. In this case, ${\rm H_2}$ is the more 
abundant if the temperature is the lower. 

Figure 1 shows the change of cooling and heating rates per unit volume 
($\Lambda$ and $\Gamma$) with the distance from the central massive 
star for the typical cloud ($n=10~{\rm cm}^{-3}$,  
$T=3000$ K). Here we have assumed the existence of 
one O5 star, whose mass is $\sim 40 ~M_{\sun}$ and 
luminosity of the LW bands is $\sim 10^{24}$ ergs s$^{-1}$ Hz$^{-1}$
\footnote{Note that although the total luminosity of a star 
depends strongly on the mass, the dependence of the luminosity 
in the LW bands depends is rather weak.}, 
at the center of the cloud, and also assumed $n$,  
$T$ and $x_e$ are constant in space, for simplicity. 
At the inner region where LW band radiation is not shielded, 
OI and CII line cooling is the dominant cooling process, and hence 
the cooling rate is approximately proportional to $z$. 
On the other hand, at the outer region where LW band radiation is shielded, 
H$_2$ line cooling becomes dominant for metal poor clouds. 

We calculate the cooling time 
\be
t_{\rm cool}=\frac{(3/2) n k T}{\Lambda_{\rm eff}}, 
\ee
where $n$ and $T$ are the number density and the temperature of the 
cloud and  
$\Lambda_{\rm eff} ~(\equiv \Lambda - \Gamma)$ is the effective 
cooling rate. 
Figure 2 shows the change of the cooling time with the distance from 
the center. 

If metallicity is slightly high ($z/z_\odot \gsim 10^{-2}$), 
the cooling time is shorter than the free-fall time 
$t_{\rm ff}~(\equiv (\frac{3 \pi}{32 G \mu m_{\rm H} n})^{1/2} 
\simeq 1.5 \times 10^7$ yr $(n/10~{\rm cm}^{-3})^{-1/2})$ at the 
all region. 
Here $G$ is the gravitational constant, $\mu$ is the mean atomic weight
and  $m_{\rm H}$ is the hydrogen mass. 
Considering HI region, almost all Hydrogen is atomic and 
$\mu \simeq 1.4$. 
On the contrary, if metallicity is very low, 
$t_{\rm ff} < t_{\rm cool}$ at the inner region and 
$t_{\rm ff} > t_{\rm cool}$ at the outer region.  
For the low density ($n \sim 1 {\rm cm}^{-3}$) and high metallicity  
($z/z_\odot \sim 1$) case, there exists no cool region where net 
cooling is negative (heating occurs) at the inner region. 

For all cases shown in Fig. 2, $t_{\rm cool} < t_{\rm ff}$ is achieved at 
the region outer than a certain distance from the center, and 
we call this transition radius as the cooling radius ($r_{\rm cool}$). 
However, in the cases of $z/z_\odot=10^{-1}$ and 1, 
the Str\"{o}mgren radius, $r_{\rm S}$, is 
larger than $r_{\rm cool}$. 
If $r < r_{\rm S}$, the gas is fully ionized and the temperature becomes 
high, star formation is strongly suppressed. 
Thus, we take as  
actual $r_{\rm cool}$ the largest between $r_{\rm cool}$ and $r_{\rm S}$. 
In a region $r \leq r_{\rm cool}$, next generation stars are hardly 
formed before the death of the central star, 
since cooling is inefficient ($t_{\rm cool} > t_{\rm ff}$).  
Thus, we consider this region as a influence region. 
For the very lower metallicity case, $z/z_\odot \lsim 10^{-2.5}$, 
$r_{\rm cool}$ depends on $z$ very weakly, since main coolant becomes 
H$_2$ at the outer region. 
If we adopt $x_{\rm H^+} = 0$, $n_{\rm H_2}$ depends a little on $z$  
so that the influence radius depends a little on $z$. 
But the overall tendency does not change from the case of 
$x_{\rm H^+}=10^{-4}$. 

The Str\"{o}mgren radius is
\begin{eqnarray}
r_{\rm S} 
\simeq 24 {\rm pc} ~\left(\frac{n_{\rm H}}{10~{\rm cm^{-3}}}\right)^{-2/3} 
\left(\frac{Q_{\ast}}{5.1 \times 10^{49} {\rm s^{-1}}}\right)^{1/3}, 
\end{eqnarray} 
where $Q_{\ast}$ is the flux of ionizing photons by a OB star and
$Q_{\ast} \simeq 5.1 \times 10^{49} {\rm s^{-1}}$ for an O5 star. 
The mass within $r_{\rm S}$ is 
\be
M_{\rm s} \simeq 2 \times 10^4 M_\odot 
~\left(\frac{n_{\rm H}}{10~{\rm cm^{-3}}}\right)^{-1}
\left(\frac{Q_{\ast}}{5.1 \times 10^{49} {\rm s^{-1}}}\right).  
\ee
$M_{\rm s}$ is somewhat smaller than the typical cloud Jeans mass 
(see Figs. 3-5). 
Note that  $r_{\rm S}$ and $Q_*$ depend strongly on 
the mass of the central star. 
Thus, if we consider a less massive 
central star ($M \simeq 10 M_\odot$), $r_s$ becomes much smaller but 
other features of these figures hardly change. 

For some cases, especially for the high temperature and 
very low metallicity case, 
$t_{\rm cool}$ can not be shorter than $t_{\rm ff}$ for  
any distance case because of the insufficiency of $x_{\rm H_2}$. 
In this case, $r_{\rm cool}$ is estimated to be $\infty$. 

\section{Star Formation Regulation in Low Metallicity Clouds}

To evaluate the strength of the star formation regulation, 
we calculate the ratio of the cooling radius to the Jeans length,  
$r_{\rm cool}/ r_{\rm J}$, in the $n$-$T$ plane for the clouds with 
various metallicity ($0 \leq z/z_\odot \leq 1$). 
Here, $r_{\rm J} \equiv ({ \pi k T \over G \mu m_{\rm H} \rho} )^{1/2} 
\simeq 2.1 \times 10^2 {\rm pc} 
(T/3000 ~{\rm K})^{1/2} (n/10~{\rm cm^{-3}})^{-1/2}$. 
As shown in Fig. 3, for the case of $z/z_\odot=10^{-2}$, 
$n$-$T$ plane is divided into two regions. 
In the region with higher density and/or lower temperature, 
$r_{\rm cool}/ r_{\rm J}$ is smaller than unity. 
In other word, influence radius of a massive star is 
smaller than the typical scale of the cloud ($r_{\rm J}$). 
Thus, the regulation is considered to be ineffective. 
In the other region, $r_{\rm cool}/ r_{\rm J} > 1$ and 
the regulation works well. 
For dense clouds, in the right lightly shaded region, 
$t_{\rm cool} < t_{\rm ff}$ for the whole HI region. 
In this case, $r_{\rm cool} = r_{\rm S} \ll r_{\rm J}$. 
For low density and high temperature clouds, in the upper left 
shaded region,  
$t_{\rm cool} > t_{\rm ff}$ for the whole HI region. In this case, 
$r_{\rm cool} \gg r_{\rm J}$. 

Figure 4 shows 
$r_{\rm cool}/ r_{\rm J}$ on the $n$-$T$ plane, for the case of 
$z/z_\odot=10^{-1.5}$. 
In this case $r_{\rm cool}/ r_{\rm J} < 1$ for almost the whole region. 
Thus, star formation regulation by UV radiation is inefficient.  
On the contrary, as shown in Fig. 5, for 
the case of $z/z_\odot=10^{-2.5}$, $r_{\rm cool}/ r_{\rm J} > 1$ 
for almost the whole region and 
the regulation works very well. 
The transition occurs when the metallicity is  
$10^{-2.5} \lsim z/z_\odot \lsim 10^{-1.5}$, 
and the typical metallicity when the transition occurs is 
estimated as $z/z_\odot \simeq 10^{-2}$. 

For the extremely low-metallicity case 
($z/z_\odot \lsim 10^{-4}$), the effect of heavy elements on the 
thermal process is almost negligible. 
Thus, considering the thermal process, it is difficult to 
distinguish a cloud with $z/z_\odot \lsim 10^{-4}$ from 
a cloud with $z/z_\odot = 0$. 

\section{Discussion}

For the primordial gas clouds, if we consider the self-shielding effect,  
the mass within the region of influence is obtained as (\cite{ON99}) 
\be
M^{\rm (inf)} 
=  8 \times 10^6 M_{\sun} \left(\frac{x_{\rm e}}{10^{-4}}\right)^{-1}
\left(\frac{L_{\rm LW}}{10^{24} {\rm ergs~s^{-1} Hz^{-1}}}\right)
\left(\frac{T}{3 \times 10^3 {\rm K}}\right)^{-1} 
\left(\frac{n}{10 {\rm cm^{-3}}}\right)^{-1}.  
\ee 
This mass is larger than the typical cloud mass (see Fig. 3-5). 
On the contrary, for the low metallicity gas clouds, the strength of 
the self-regulation changes with the metallicity.  

As shown in the previous section, in the case of 
very low metallicity clouds 
($z/z_\odot \lsim 10^{-2.5}$), star formation regulation by UV radiation 
is very effective. 
In this case, star formation rate is very low, since only one massive star 
can stop the evolution of the whole host cloud.  
If SNe do not disrupt the gas binding, 
the host cloud is polluted by heavy elements little by little and 
the following continuous star formation is possible. 
Even in the case that the host cloud is disrupted by SN explosions, 
if the remnant gas does not escape from the host pregalactic object, 
next generation clouds, which are slightly polluted, will be formed 
and subsequent star formation will follows.  
After metallicity becomes high enough ($z/z_\odot \gsim 10^{-2}$), 
effective star formation can start. 
Thus, the lower limit of 
metallicity of luminous objects is roughly estimated at 
$z/z_\odot \sim 10^{-2}$. 
By the way, a SN release several $\times M_\odot$ of heavy elements 
and the typical cloud mass ($\sim \rho r_{\rm J}^3$) is 
$\sim 10^6 ~M_\odot$. 
Thus, if we consider that the cloud mass is $10^6 ~M_\odot$ and 
the mass of the heavy elements released by one SN is $4 M_\odot$, 
the metallicity increase per one SN is 
$\sim 4 \times 10^{-6}$ and 
the change of the metallicity is $\sim 2 \times 10^{-4} z_\odot$, 
and hence 
before efficient star formation begins, about 50 cycles 
of star formation and SN explosion is required. 
This implies that at the effective star formation epoch, 
there should exist some amount of heavy elements and dust 
and they are scattered well. Then, it is expected that it is 
hard for Ly $\alpha$ photons to escape from the clouds. 
Moreover, the inefficiency of star formation in low metallicity 
clouds may explain the G dwarf problem (e.g., \cite{RM96}) 
and may also result in 
that the reionization epoch of the universe may become later than the 
previous studies (e.g., \cite{FK94,HL97,GO97}). 

The lifetime of very massive stars is $\sim 3 \times 10^6$ yr. 
However, as noted above, the mass dependence of the luminosity 
in the LW bands is rather weak, and hence even if the mass of 
the central star is smaller ($\lsim 10~M_\odot$), 
in this case the lifetime of the star is longer, 
star formation regulation can be efficient. 
Moreover, these less massive stars may not evolve into type II SNe. 
Thus, after the death of the first OB star, star formation could occur
somewhere in the cloud, and another OB star could form successively.
Therefore, star formation regulation becomes often serious 
even for clouds whose $t_{\rm ff}$ is longer than  $\sim 3 \times 10^6$ yr. 

Lin \& Murray (1992) considered the regulation only via 
photoionization. 
In such a case, the affected region by an OB star becomes the region 
within a Str\"{o}mgren sphere. 
For the very lower metallicity case, $r_{\rm S}$ is somewhat smaller than 
$r_{\rm cool}$. 
However, if metallicity is fairly high $z/z_\odot \gsim 10^{-1.5}$, 
$r_{\rm cool}$ becomes $r_{\rm S}$ for almost the whole region in $n$-$T$ 
plane (see lightly shaded region in Fig. 4). 
Thus, for the higher metallicity case 
($z/z_\odot \gsim 10^{-1.5}$), it is good approximation to consider that 
star formation regulation occurs 
only via photoionization, but in this case influence region is 
much smaller than the host cloud hence the regulation is not effective. 

Recently, it has been shown that earlier phase of the chemical evolution 
of the Galaxy can be explained by the models with the assumption that 
SN-induced star formation is the only star formation process 
(e.g., \cite{Tet99,IW99}).  
As shown above, when the metallicity is lower ($z/z_\odot \lsim 10^{-2}$), 
star formation can occur only after previous massive stars have died. 
If these massive stars dye with SN explosion, 
we can consider that SN-induced star formation is the only 
star formation process, since stars can form only after SN 
explosion. 

\acknowledgments 
We would like to thank H. Sato and anonymous referee 
for valuable comments. 
This work is supported in part by the 
Grant-in-Aid for Scientific Research on Priority Areas 
(No. 10147105) of the Ministry of Education,
Science, Sports and Culture of Japan (RN). 

\newpage

\newpage
\begin{figure}
\plotone{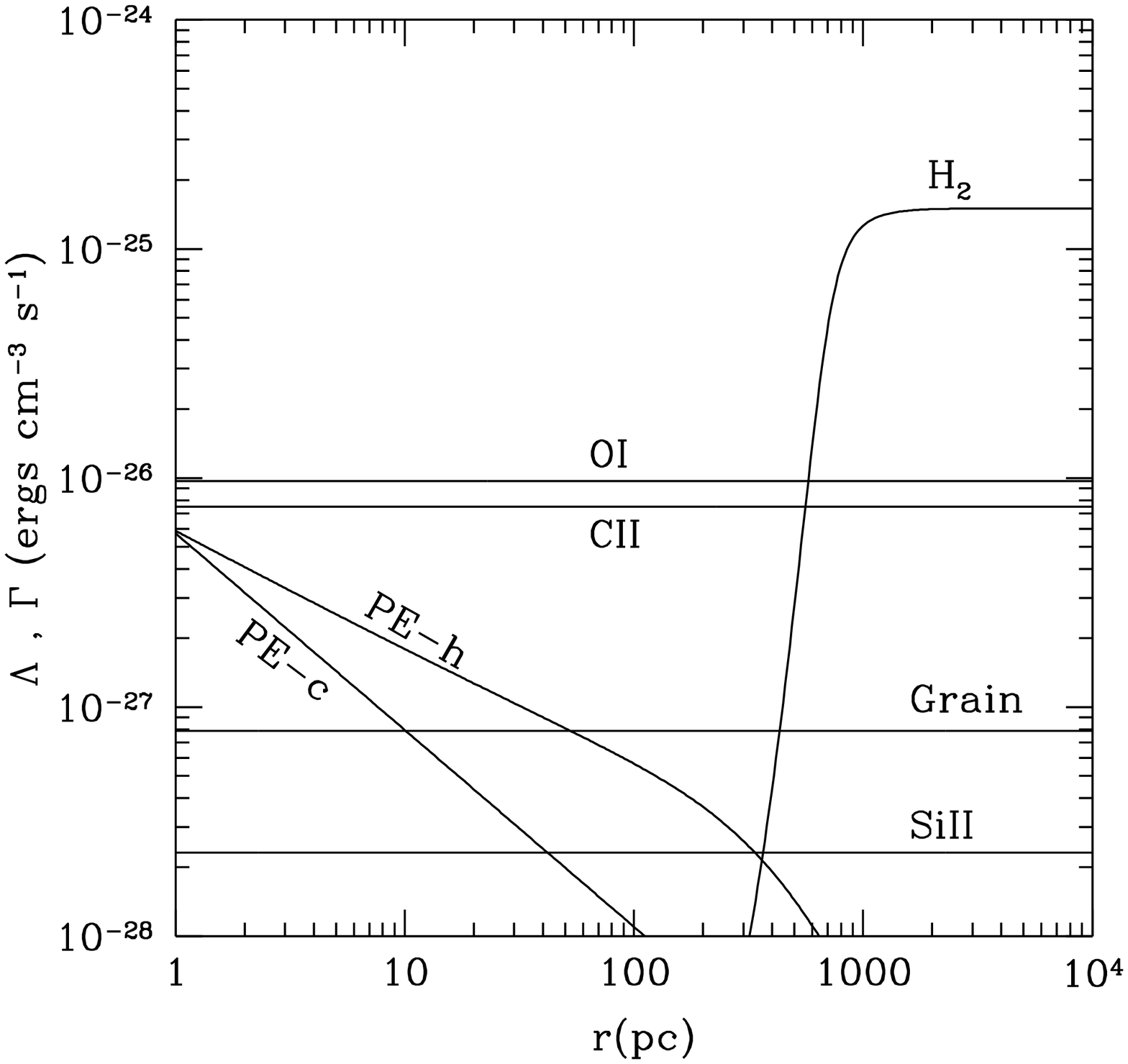}
\caption[dummy]
{Change of the cooling and heating rates per unit volume 
($\Lambda$ and $\Gamma$) with the distance from the central massive 
star for the typical cloud ($n=10~{\rm cm}^{-3}$,  
$T=3000$ K and $z = 10^{-2} z_\odot$). 
The central star is assumed to be one O5 star. 
PE-h and PE-c in the figure denote the grain photoelectric heating 
and cooling rates, respectively. 
The others denote radiative cooling rates of 
H$_2$, OI, CII, SiII and Grain, respectively. 
}
\label{fig1}
\end{figure}
\begin{figure}
\plotone{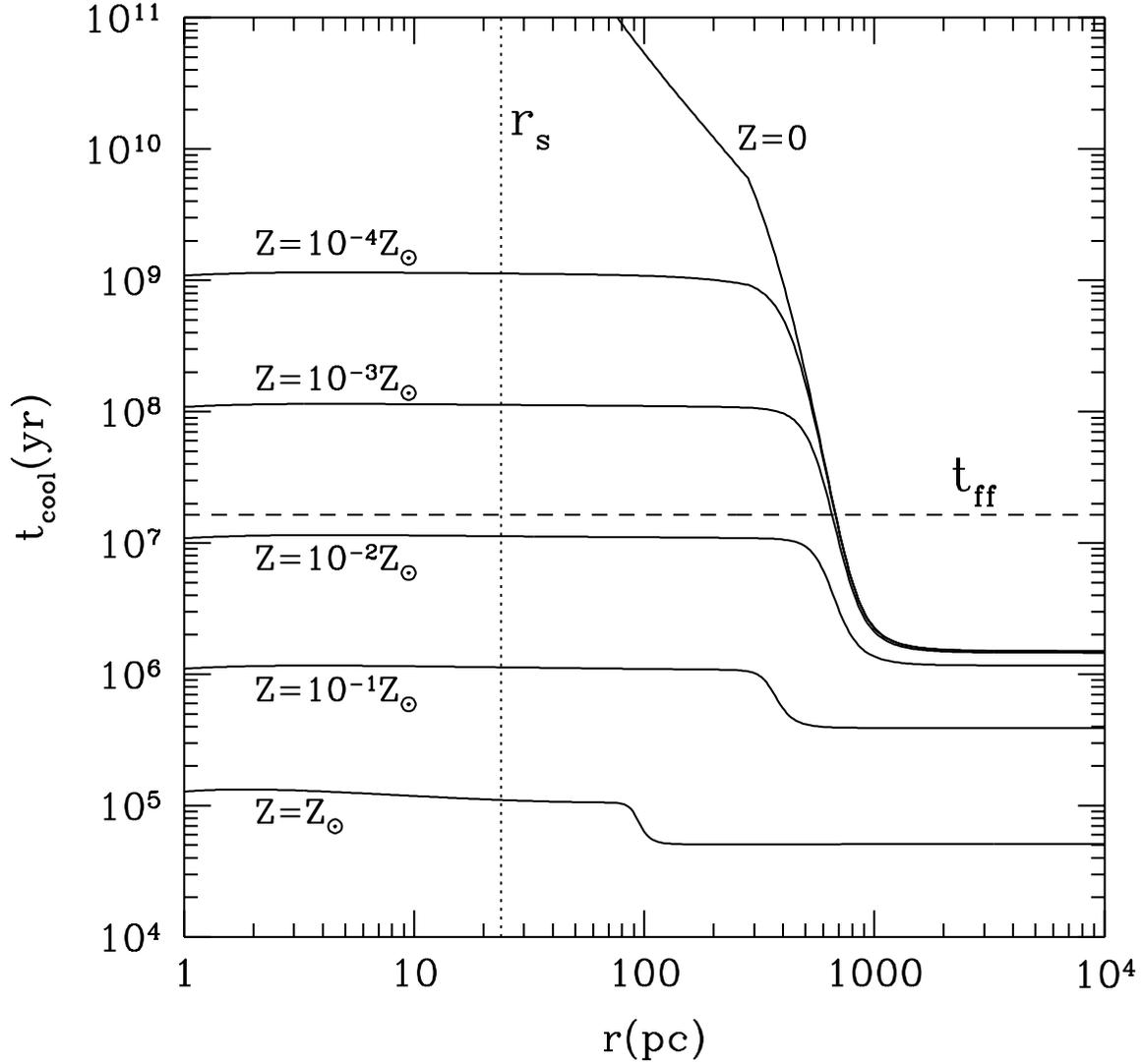}
\caption[dummy]
{Cooling time $t_{\rm cool}$ at the various distance from 
the center for the typical cloud ($n=10~{\rm cm}^{-3}$ and  
$T=3000$ K).  The metallicity is $z/z_\odot = 0$, $10^{-4}$,  $10^{-3}$, 
$10^{-2}$, $10^{-1}$ and 1.  
The dotted line denotes the Str\"{o}mgren radius, $r_{\rm S}$ 
for a O5 star. 
The dashed line denotes the free-fall time, $t_{\rm ff}$. 
}
\label{fig2}
\end{figure}
\begin{figure}
\plotone{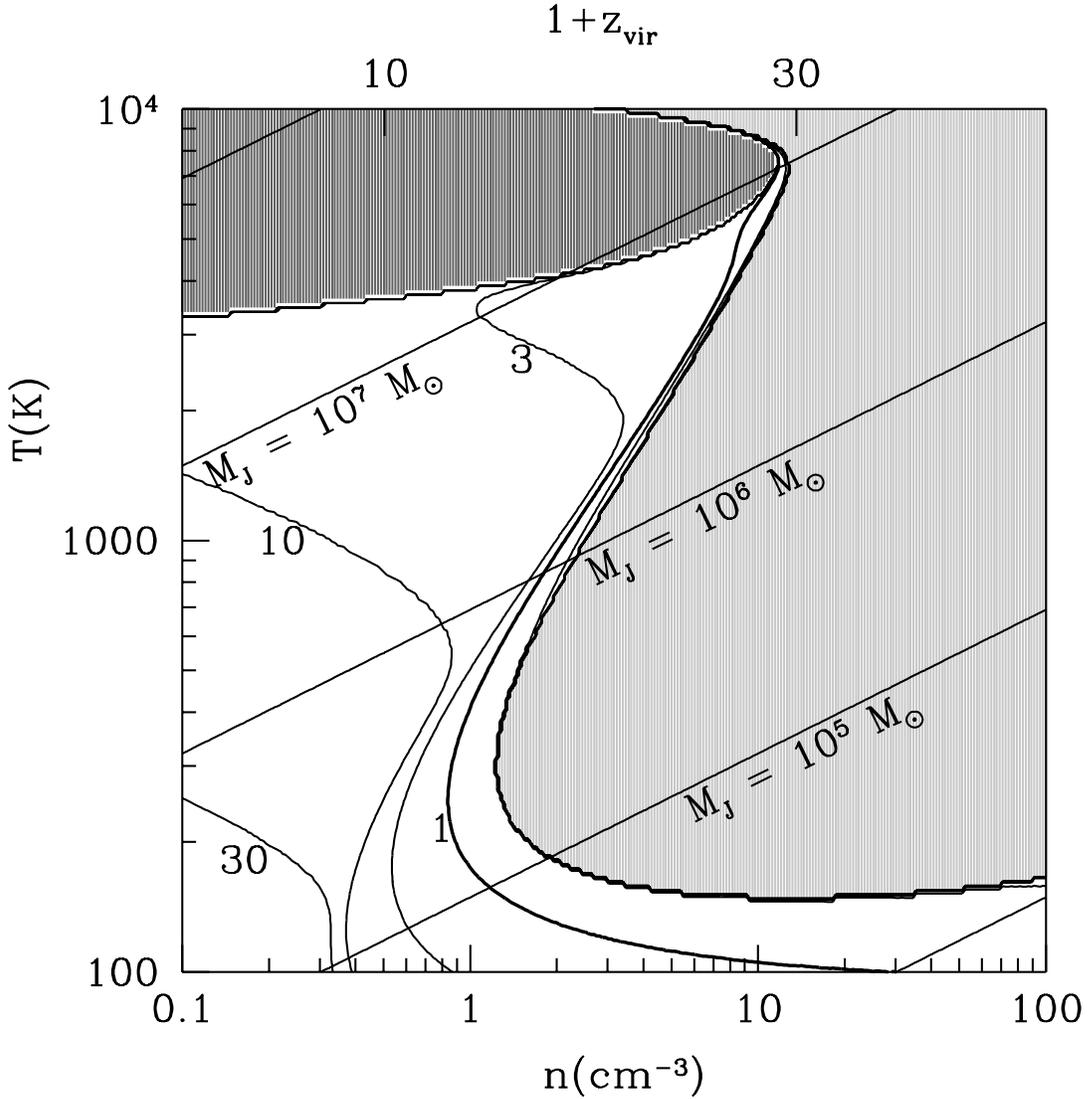}
\caption[dummy]
{The ratio of the cooling radius to the Jeans length 
$r_{\rm cool}/ r_{\rm J}$ for the case of $z/z_\odot=10^{-2}$.
The numbers in the figure denote the value of $r_{\rm cool}/ r_{\rm J}$ 
of the contour. For dense cloud, 
in the right lightly shaded region, $t_{\rm cool} < t_{\rm ff}$ 
for the whole HI region. 
In this case, $r_{\rm cool} = r_{\rm S} \ll r_{\rm J}$. 
For low density and high temperature cloud, 
in the upper left shaded region,  $t_{\rm cool} > t_{\rm ff}$ 
for the whole HI region. In this case, 
$r_{\rm cool} \gg r_{\rm J}$. 
The values of the Jeans mass are also shown. 
For reference, considering the cosmological objects, 
related virialized redshift is also shown on the upper axis 
for the flat universe with $\Omega_b = 1$. 
}
\label{fig3}
\end{figure}
\begin{figure}
\plotone{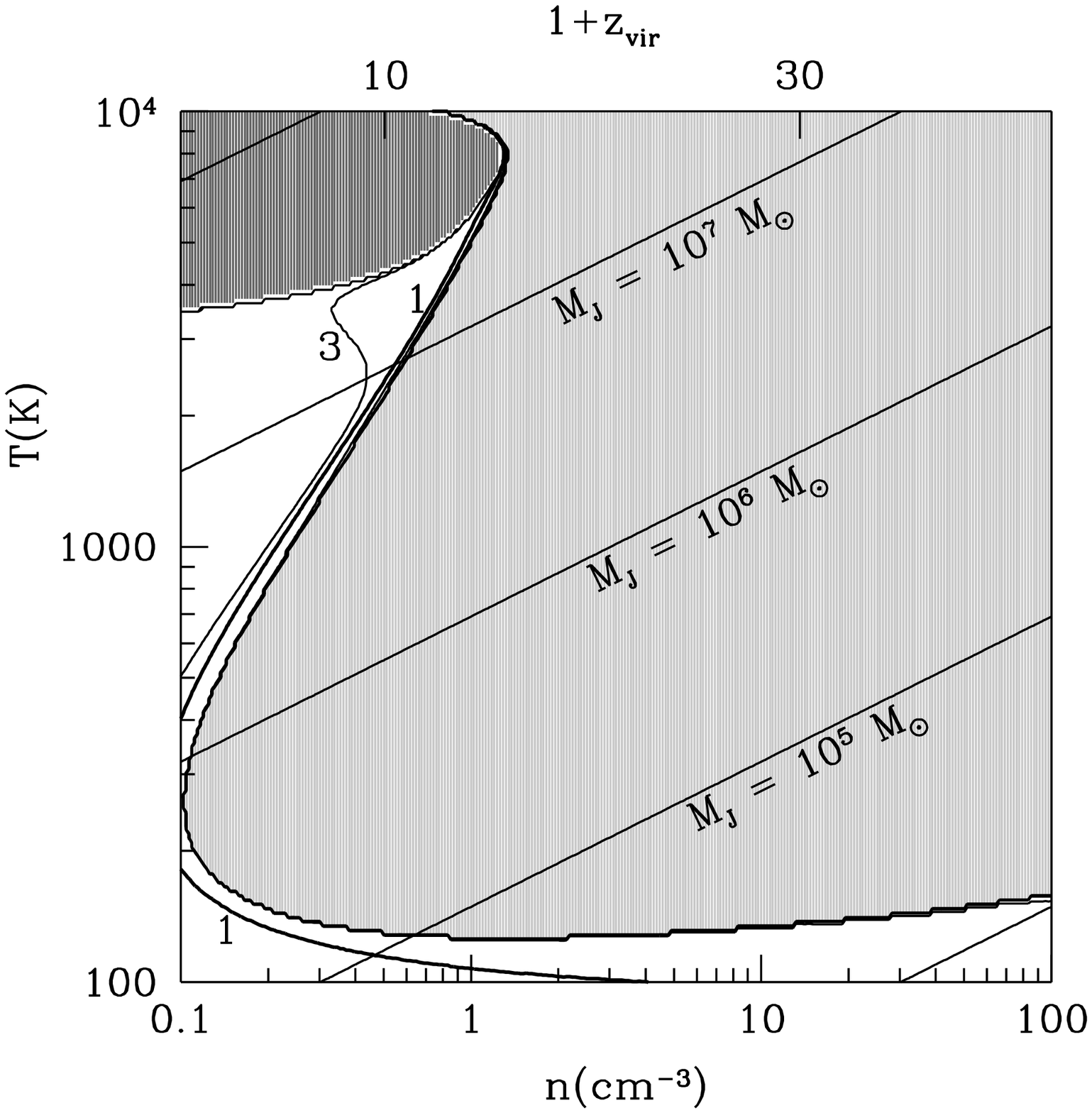}
\caption[dummy]
{The same figure of Fig. 3 but for the case of 
$z/z_\odot=10^{-1.5}$.
}
\label{fig4}
\end{figure}

\begin{figure}
\plotone{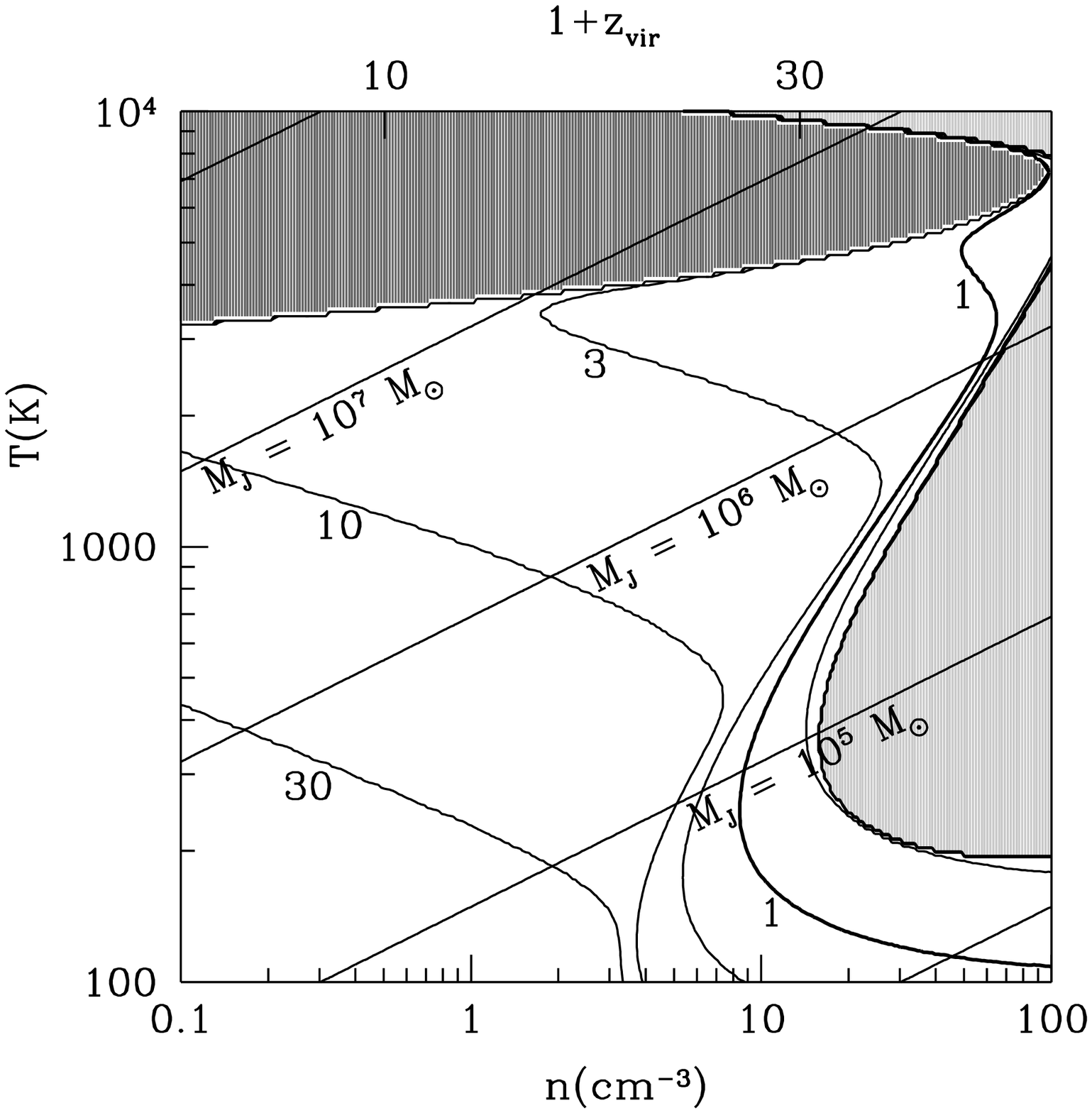}
\caption[dummy]
{The same figure of Fig. 3 but for the case of 
$z/z_\odot=10^{-2.5}$.
}
\label{fig5}
\end{figure}
\end{document}